\RequirePackage[running]{lineno}
\pdfoutput=1
\documentclass[aps,prl,nofootinbib,twocolumn,showpacs,superscriptaddress,letterpaper,amsmath,amssymb]{revtex4}
\usepackage[hyperindex,breaklinks]{hyperref}
\usepackage{graphicx} 
\usepackage{dcolumn}  
\usepackage{bm}       
\usepackage{feynmf}
\usepackage{slashed}
\usepackage{multirow}
\usepackage{placeins}

\usepackage{atlasphysics} 
\usepackage{subfigure}
\usepackage{units}

\def\gsim{\lower0.5ex\hbox{$\:\buildrel >\over\sim\:$}}
\def\lsim{\lower0.5ex\hbox{$\:\buildrel <\over\sim\:$}}

\def\metv{$\textbf{\textit{E}}_{\rm T}^{\rm miss}$}

\def\MET{E_{\rm T}^{\rm miss}}

\DeclareGraphicsExtensions{.pdf,.png,.jpg,.jpeg,.eps}

\newcommand{\rom}[1]{\uppercase\expandafter{\romannumeral #1\relax}}

\newcommand{\pythia}{\textsc{Pythia}}
\newcommand{\pythiaeight}{\textsc{Pythia8}}

\newcommand{\madgraph}{\textsc{Madgraph}}

\usepackage{preprintcover} 

\PreprintCoverPaperTitle{Search for Dark Matter in Events with Missing Transverse Momentum and a Higgs Boson Decaying to Two Photons in $pp$ Collisions at $\sqrt{s}=8$~TeV with the ATLAS Detector}

\PreprintIdNumber{CERN-PH-EP-2015-120}  

\PreprintCoverAbstract{Results of a search for new phenomena in events with large missing transverse momentum and a Higgs boson decaying to two photons are reported.  Data from proton--proton collisions at a center-of-mass energy of 8 TeV and corresponding to an integrated luminosity of \mbox{20.3 fb$^{-1}$} have been collected with the ATLAS detector at the LHC.  The observed data are well described by the expected Standard Model backgrounds.  Upper limits on the cross section of events with large missing transverse momentum and a Higgs boson candidate are also placed. Exclusion limits are presented for models of physics beyond the Standard Model featuring dark-matter candidates.}

\PreprintJournalName{Phys.~Rev.~Lett.}

\begin{document}

\thispagestyle{empty}
\vspace*{-3.5cm}

\vspace{0.5in}

\title{Search for Dark Matter in Events with Missing Transverse Momentum and a Higgs Boson Decaying to Two Photons in $pp$ Collisions at $\sqrt{s}=8$~TeV with the ATLAS Detector}

\begin{center}
\begin{abstract}
Results of a search for new phenomena in events with large missing transverse momentum and a Higgs boson decaying to two photons are reported.  Data from proton--proton collisions at a center-of-mass energy of 8 TeV and corresponding to an integrated luminosity of \mbox{20.3 fb$^{-1}$} have been collected with the ATLAS detector at the LHC.  The observed data are well described by the expected Standard Model backgrounds.  Upper limits on the cross section of events with large missing transverse momentum and a Higgs boson candidate are also placed. Exclusion limits are presented for models of physics beyond the Standard Model featuring dark-matter candidates.
\end{abstract}
\end{center}

\author{ATLAS Collaboration}


\date{ \today}
\maketitle


Although the existence of dark matter (DM) is well established, nearly nothing is known of its underlying particle nature~\cite{Bertone:2004pz}. Many DM candidates have been proposed, and attempts made to connect them to physics beyond the Standard Model (SM) at the scale of electroweak symmetry breaking~\cite{Jungman:1995df} that would naturally accommodate the observed relic density~\cite{Scherrer:1985zt}.

\begin{figure}
\includegraphics[width=4cm]{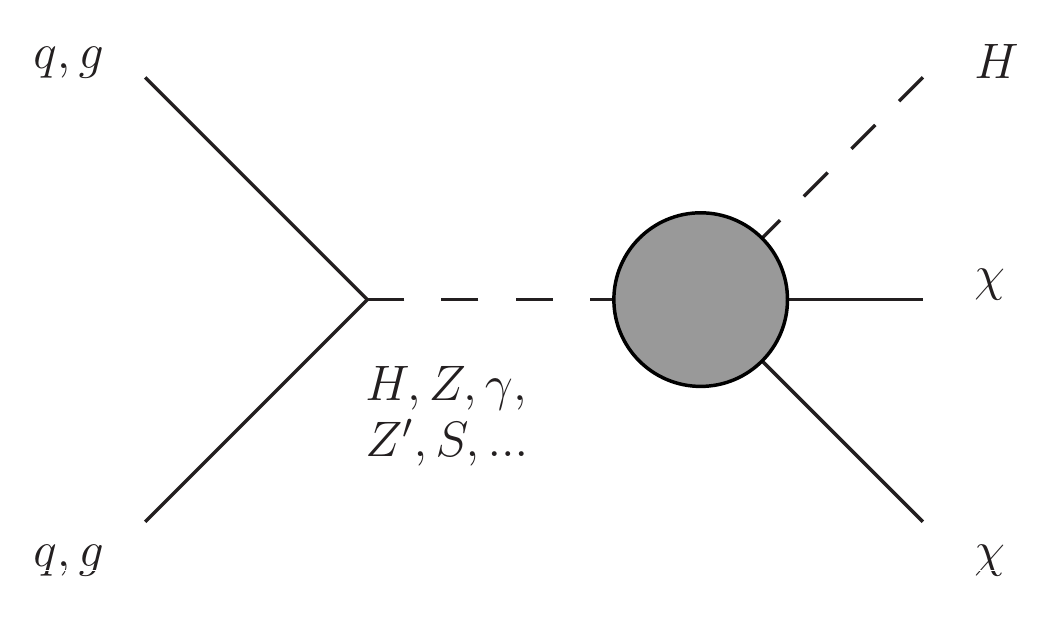}
\caption{Schematic diagram for production of DM particles $\chi$ in association with a Higgs boson in $pp$ collisions, mediated by electroweak bosons ($H,Z,\gamma$) or new mediator particles such as a $Z^\prime$ or scalar singlet $S$. The gray circle denotes an effective interaction between DM, the Higgs boson, and other states.}
\label{fig:diag}
\end{figure}

Collider searches for weakly interacting dark matter rely on the inferred observation of missing transverse momentum~\cite{coord} $\MET$ recoiling against a visible final-state object $X$, which may be a hadronic jet~\cite{Chatrchyan:2012me,Aad:2011xw}, photon ($\gamma$)~\cite{Aad:2012fw,Chatrchyan:2012tea}, or $W/Z$ boson~\cite{Carpenter:2012rg,Aad:2014vka,Aad:2013oja}. The discovery of a Higgs boson~\cite{Aad:2012tfa,Chatrchyan:2012ufa} ($H$) creates a new opportunity to search for beyond-the-SM (BSM) physics giving rise to $H+\MET$ signatures~\cite{Carpenter:2013xra,Petrov:2013nia}. In contrast to the aforementioned probes, the visible $H$ boson is unlikely to be radiated from an initial-state quark or gluon. This has the important consequence that the $H+\MET$ signature directly probes the structure of the effective DM--SM coupling; see Fig.~\ref{fig:diag}.

If the mass of the DM particle is less than half of the Higgs boson mass $m_H$, the Higgs boson may decay directly to DM.  Such decays have been searched for using LHC data, and null results provide powerful constraints on the invisible branching ratio of the Higgs boson in several different production modes including $WH$ or $ZH$~\cite{Aad:2013oja,Aad:2014iia,Aad:2015uga}, and $qqH$~\cite{Chatrchyan:2014tja,ATLAS-CONF-2015-004}. However, the mass of the DM particle may be larger than $m_H/2$, in which case these searches are not sensitive, and approaches such as analysis of $H+\MET$ events are required.

Two approaches are commonly used to model generic processes yielding a final state with a particle $X$ recoiling against a system of noninteracting particles.  One option is to use nonrenormalizable operators in an effective field theory (EFT), which is agnostic about the details of the theory at energies beyond the experimental sensitivity. Alternatively, simplified models that explicitly include the particles at higher masses can be used. The EFT approach is more model-independent, but is not valid when the typical momentum transfer approaches the scale of the high-mass particles that have been integrated out.  Simplified models do not suffer from these concerns, but include more assumptions by design and are therefore less generic.  The two approaches are thus complementary and both are considered here.

In this Letter, results are reported from a search for $H+\MET$ events in data collected by the ATLAS detector from $pp$ collisions with center-of-mass energy $\sqrt{s} = 8~\TeV$ and corresponding to an integrated luminosity of 20.3 fb$^{-1}$, produced by the Large Hadron Collider. The $H\rightarrow\gamma\gamma$ decay mode is used exclusively, as the small branching ratio is mitigated by the 
distinct diphoton resonance signature 
and the low expected number of background events with significant $\MET$~\cite{Carpenter:2013xra}.  ATLAS measured previously the differential cross section of $H\rightarrow\gamma\gamma$ production with respect to several kinematic quantities~\cite{Aad:2014lwa}, including $\MET$; the search reported here uses a subset of those data optimized for sensitivity to production of dark matter in association with a Higgs boson.


The ATLAS detector~\cite{ATLAS} is a multipurpose particle
physics experiment with a forward-backward symmetric
cylindrical geometry and nearly 4$\pi$ coverage in solid angle.
Events were selected using a trigger that requires two photons, with leading (subleading) \et\ $>35~(25)$ GeV.


A photon is reconstructed as a cluster of energy with $|\eta|<2.37$ deposited in the electromagnetic calorimeter,  excluding the poorly instrumented region $\eta \in [1.37,1.56]$.  Clusters without matching tracks are classified as unconverted photon candidates.
The photon energy is corrected by applying an energy calibration derived from
$Z \rightarrow e^+ e^-$ decays in data and cross-checked with $J/\psi \rightarrow e^+ e^-$ and $Z\rightarrow\ell\ell\gamma$ decays in data~\cite{Aad:2014nim}.
Identification requirements are applied in order to reduce the contamination dominantly from $\pi^0$ or other neutral hadrons decaying to two photons.  The photon identification is based on the profile of the energy deposit in the first  and second layers of the electromagnetic calorimeter. Photons have to satisfy the `tight' identification criteria of Ref.~\cite{Aad:2014eha}. They are also required to be isolated, i.e.  the energy in the calorimeters  in a cone of size $\Delta R = \sqrt{(\Delta \eta)^2 + (\Delta \phi)^2} = 0.4$ around the cluster barycenter, excluding the energy associated with the photon cluster, is required to be less than $6\GeV$. This in-cone energy is corrected for the leakage of the photon energy and for the effects of
multiple $pp$  interactions in the same or neighboring bunch crossings  
superimposed on the hard physics process (referred to as  pileup interactions)  \cite{PhIsol}. Finally, for each photon the scalar sum of the transverse momenta \pT\ of tracks originating from the diphoton vertex with \mbox{$\pT > 1$ GeV} and \mbox{$\Delta R(\textrm{track,cluster})<0.2$} must be less than 2.6~GeV. The diphoton production vertex is selected from the reconstructed collision vertices using a neural-network algorithm as described in Ref.~\cite{Aad:2014eha}.

The momentum imbalance in the transverse plane is obtained from  the negative vector sum of the reconstructed and calibrated electrons, muons, photons and jets and  is referred to as missing transverse momentum \metv. The symbol $\MET$ is used for its magnitude. Calorimeter energy deposits are associated with a reconstructed and identified high-$\pT$ object in a specific order: photons with \pT\ $>10\GeV$, electrons with \pT\ $>10\GeV$,  and  jets with \mbox{$\pT > 20\GeV$}. Deposits not associated with any such objects are also taken into account in the \metv calculation \cite{ATLAS-CONF-2013-082} using an energy-flow algorithm that considers calorimeter energy deposits as well as inner-detector tracks \cite{Aad:2012re}. The energy resolution is typically 11\% near the threshold at 100 GeV for the considered signal scenarios. 


Quality requirements are applied to photon candidates in order to reject those arising from instrumental problems. In addition, quality requirements are applied in order to remove jets arising from detector noise or out-of-time energy deposits in the calorimeter from cosmic rays or other noncollision processes \cite{Aad:2014bia}. 

Selected events are required to have a Higgs boson candidate consisting of two photons with diphoton invariant mass $m_{\gamma\gamma} \in [105,160]$ GeV with transverse momenta satisfying leading (subleading) $\pT^{\gamma} > 0.35 (0.25) m_{\gamma\gamma}$. In addition, large  missing transverse momentum is required, \mbox{$\MET > 90\GeV$}, as well as large transverse momentum of the $\gamma\gamma$ system, \mbox{$\pT^{\gamma\gamma} > 90$ GeV} in order to suppress background events where $\MET$ is caused by mismeasurement of the energies of identified physics objects.  These selection requirements were derived by optimizing the expected upper limits on $H+\MET$ production for the set of models described below.

Contributions to the $\gamma\gamma+\MET$ sample from SM processes include those that produce a Higgs boson in association with undetected particles (predominantly $ZH$ with $Z\rightarrow\nu\bar{\nu}$ and $WH$ with $W\rightarrow\ell\nu$) as well as nonresonant diphoton production ($\gamma\gamma$, $W\gamma\gamma$, $Z\gamma\gamma$), $W\gamma$ and $Z\gamma$ production where an electron is misidentified as a photon, and photon+jet production in which the jet is misidentified as a photon. 


Samples of simulated events are used in order to measure the efficiency of the selection for dark-matter models, as well as to estimate the contribution of SM $H+\MET$ processes.  Contributions from other background processes are estimated from $m_{\gamma\gamma}$ sidebands in the data.

Following the notation of Ref.~\cite{Carpenter:2013xra}, a set of EFT models are considered in which the effective operator Lagrangian term can be written as $|\chi|^2|H|^2$, $\bar{\chi}i\gamma_5\chi|H|^2$, $\chi^\dagger \partial^\mu \chi  H^\dagger D_\mu H$, or $\bar \chi \gamma^\mu \chi  B_{\mu\nu} H^\dagger D^\nu H$, where the DM field $\chi$ is a scalar in the first case and a fermion in the remaining cases and $B_{\mu\nu}$ is the $U(1)_Y$ field strength tensor.   The interactions of SM and DM particles are described by two parameters: the DM particle mass $m_\chi$ and the suppression scale $\Lambda$ of the heavy mediator that is integrated out of the EFT. In a theory that is valid to arbitrary energies (ultraviolet complete), the contact interaction would be replaced by an interaction via an explicit mediator $V$. 

In addition, simplified models~\cite{Carpenter:2013xra} with a massive vector ($Z'$), or a scalar ($S$) intermediate boson are tested.  All $H+\MET$ DM models are generated with $\madgraph5$~\cite{Alwall:2011uj}  version 1.4.8.4, with showering and hadronization modeled with $\pythiaeight$~\cite{Sjostrand:2007gs} version 1.6.5 using the AU2 parameter settings~\cite{ATL-PHYS-PUB-2011-008}; the MSTW2008LO~\cite{Martin:2009iq} parton distribution function (PDF) set is used. Values of $m_\chi$ from 1 to 1000 GeV are considered. Production of $ZH$ and $WH$ is modeled with $\pythiaeight$ using CTEQ6L1 PDFs \cite{Pumplin:2002vw}.  Samples are normalized to cross sections for $WH$ and $ZH$ production calculated at next-to-leading order (NLO)~\cite{Han:1991ia}, and next-to-next-to-leading order (NNLO)~\cite{Brein:2003wg} in QCD, respectively, with NLO electroweak corrections~\cite{Ciccolini:2003jy} in both cases.

Differing pileup conditions as a function of the instantaneous luminosity are taken into account by overlaying simulated minimum-bias events generated with   $\pythiaeight$ onto the hard-scattering process such that the observed distribution of the average number of interactions per bunch crossing is reproduced. The simulated samples are processed with a full ATLAS detector simulation~\cite{:2010wqa} based on {\sc Geant4}~\cite{Agostinelli:2002hh} and a simulation of the trigger system. 


To distinguish contributions from processes that include $H\rightarrow \gamma\gamma$ decays from those that contribute to the continuum background, a localized excess of events is searched for in the $m_{\gamma\gamma}$ spectrum near the Higgs boson mass, $m_H=125.4$ GeV.   Probability distribution functions that describe the $H\rightarrow\gamma\gamma$ resonance or the continuum background are defined in the range 105--160 GeV as described below. The contributions from each source are then estimated using an unbinned maximum-likelihood fit to the observed $m_{\gamma\gamma}$ spectrum.

\begin{figure}
\includegraphics[width=8cm]{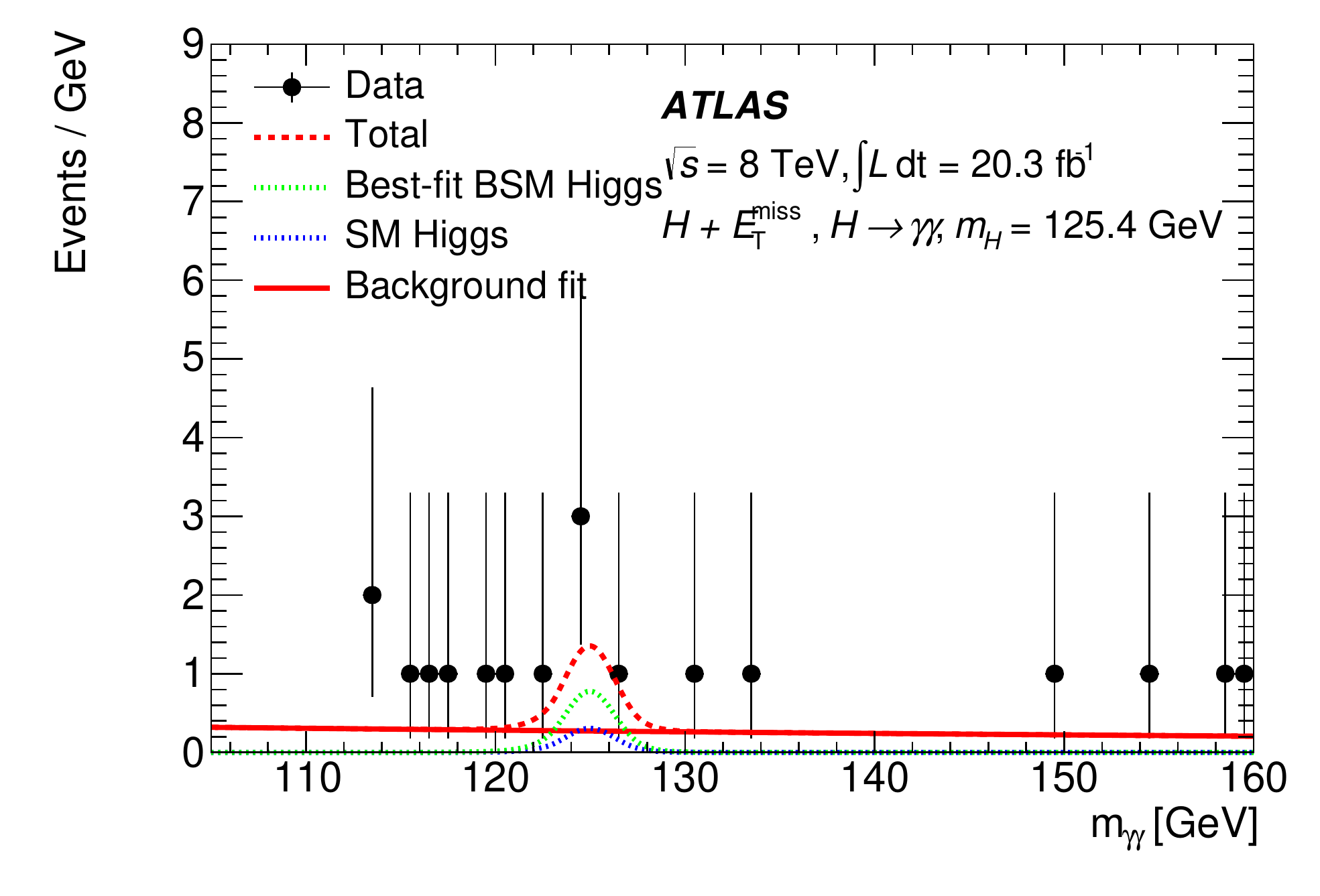}
\caption{
The best-fit background estimates to the 18 observed events are 14.2 $\pm$ 4.0 (continuum backgrounds) 1.1 $\pm$ 0.1 (SM Higgs boson backgrounds) and 2.7 $\pm$ 2.2 (BSM Higgs boson), including both statistical and systematic uncertainties. An unbinned maximum-likelihood fit to the spectrum is used to estimate the number of events from the continuum background and from $H\rightarrow \gamma\gamma$ decays; the individual components are shown as well as their sum.}
\label{fig:fit}
\end{figure}

The $m_{\gamma\gamma}$ spectra of the signal models of $H+$DM production and SM Higgs boson background processes are modeled with a double-sided Crystal Ball~\cite{Oreglia:1980cs} function; the width and peak positions are fixed to values extracted from fits to simulated samples.   An exponential function, $e^{a m_{\gamma\gamma}}$ with free parameter $a$ is used to describe the $m_{\gamma\gamma}$ distribution of the continuum background. The chosen continuum fit function is validated using simulated samples of the irreducible background processes and in three data samples adjacent to the signal region, but with relaxed requirements on $\MET$, on $\pT^{\gamma\gamma}$, or on photon identification.   Results of the fit to data in the signal region are shown in Fig.~\ref{fig:fit}.

Systematic uncertainties from various sources affect the number of SM Higgs boson events in the resonant background, the predicted shape and location of its peak, as well as the efficiency of the selection for the signal models considered.  

The uncertainty on the integrated luminosity, 2.8\%, is derived following the same methodology as that detailed in Ref.~\cite{Aad:2013ucp} using beam-separation scans.  Uncertainties on the efficiency of the photon isolation requirement, photon identification requirement, and trigger selection are measured in an inclusive SM Higgs boson sample to be 2.8\%, 2.1\%, and 0.2\%, respectively.   Uncertainties in the photon energy scale and resolution lead to respective uncertainties of 11\% and 0.3\% in the position and width of the $H\rightarrow\gamma\gamma$ peak. Additional uncertainties on the jet energy scale and resolution as well as the calibration of unclustered hadronic recoil energy contribute to uncertainty in the $\MET$, leading to 1.2\% uncertainty on the efficiency of the selection for the signal models from the $\MET$ and $p_{\textrm{T}}^{\gamma\gamma}$ requirements.  The impacts on the selection efficiency of the uncertainties on the levels of initial-state  and final-state radiation  are assessed by varying the \pythia8 parameters, as in Ref.~\cite{Aad:2014vka}; these are  found to be typically at the level of 1\%.  The total uncertainty on the selection efficiency for peaking SM Higgs backgrounds and signal models is 4.0\%.

The theoretical uncertainties on the $WH$ and $ZH$ production cross sections come from varying the renormalization and factorization scales and from uncertainties on the parton distribution functions~\cite{Lai:2010vv,Botje01,Martin:2009iq,Ball:2011mu} following the PDF4LHC prescription. The Higgs boson decay branching fractions are taken from Refs.~\cite{Djouadi:1997yw,Actis:2008ts} and their uncertainties from Refs.~\cite{Dittmaier:2011ti,Dittmaier:2012vm}. The total theoretical uncertainty on the $H+\MET$ contribution is 6\%.


The number of events observed in the data corresponds to a $1.4~\sigma$ deviation using the asymptotic formulae in Ref.~\cite{Cowan:2010js}. As the events observed do not include a statistically significant BSM component, the results are interpreted in terms of exclusions on models that would produce an excess of  $H+\MET$ events. Upper bounds, detailed below, are calculated using a one-sided profile likelihood ratio and the $CL_S$ technique~\cite{Read:2002hq,Junk:1999kv}, evaluated using the asymptotic approximation~\cite{Cowan:2010js}, which was ensured to be valid for the available number of events.

The most model-independent limits are those on the fiducial cross section of $H+\MET$ events, including SM and BSM components, $\sigma\times A$, where $\sigma$ is the cross section and $A$ is the fiducial acceptance. The latter is defined using a selection identical to that defining the signal region but applied at particle level, where \metv\ is the vector sum of the momenta of the noninteracting particles, photon isolation requirements are not applied, and a simpler requirement on photon pseudorapidity $|\eta|<2.37$ is made. The limit on $\sigma\times A$ is derived from a limit on the visible cross section $\sigma\times A \times \epsilon$, where $\epsilon$ is the reconstruction efficiency in the fiducial region. An estimate $\epsilon = 56$\% is computed using the simulated signal samples described above with no quark or gluon produced from the main interaction vertex; the efficiencies vary  across the set of models by less than 10\%.  The observed (expected)  upper limit on the fiducial cross section is 0.70 (0.43) fb at 95\% confidence level (CL).  These limits are applicable to any model that predicts $H+\MET$ events in the fiducial region and has similar reconstruction efficiency $\epsilon$.

\begin{figure}
\begin{center}
\includegraphics[width=6cm]{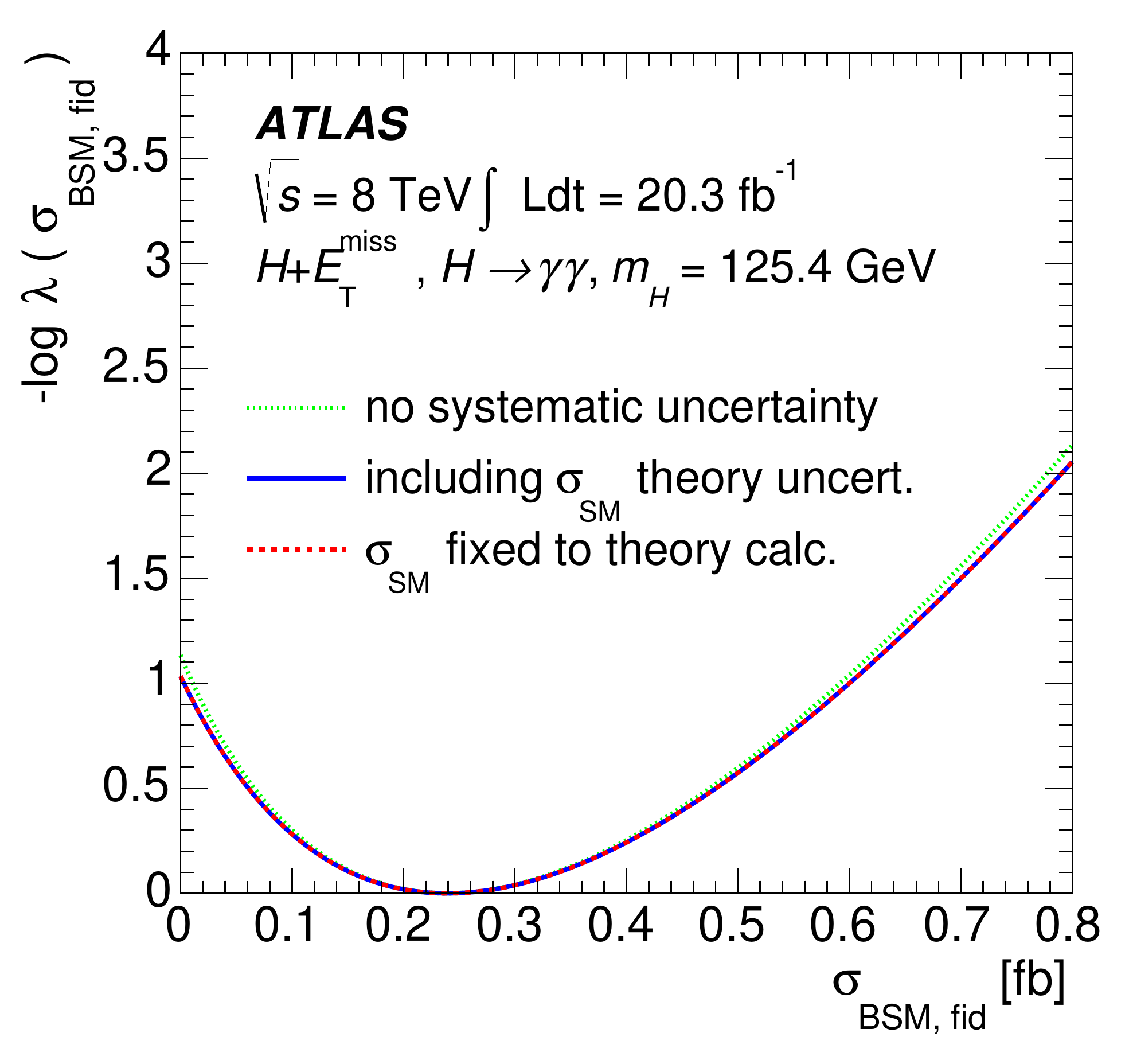}
\end{center}
\caption{Profile likelihood ratio ($\lambda$) as a function of $\sigma_{\textrm{BSM,fid}}$, the fiducial cross section for production of a BSM $H+$DM process in the $\gamma\gamma + \MET$ channel taking into account the contribution of the SM component. The solid blue likelihood curve shows that the number of events observed in the data corresponds to a $1.4~\sigma$ deviation using the asymptotic formulae in Ref.~\cite{Cowan:2010js}. The dotted green likelihood curve only includes statistical uncertainties. The dashed red likelihood curve allows for modifications of the central value and uncertainty on the SM component as described in the text.}
\label{fig:lhood}
\end{figure}

Limits on specific models of BSM $H+\MET$ production depend on the prediction of the $H+\MET$ component produced via $ZH$ or $WH$; calculations of this theoretical quantity will improve with time and may depend on the details of a specific BSM theory.  Following the proposal of Ref.~\cite{Cranmer:2013hia}, the profile likelihood ratio of the cross section for BSM $H+$DM production in the $\gamma\gamma + \MET$ channel is provided with the SM component fixed to the central value of the theoretical calculation, which allows later reinterpretation for any modified prediction and uncertainty, as shown in Fig.~\ref{fig:lhood}. This approach requires knowing how a change in the SM-like component modifies the best-fit BSM component; in this case where the SM-like and BSM components are indistinguishable, $\Delta N_{\textrm{BSM}} = -\Delta N_{\textrm{SM-like}}$.  The limits on the parameters of the specific BSM models considered in this Letter are calculated using the prediction and uncertainty for the SM component as described above.

\begin{figure}
\begin{center}
        \includegraphics[width=0.495\textwidth]{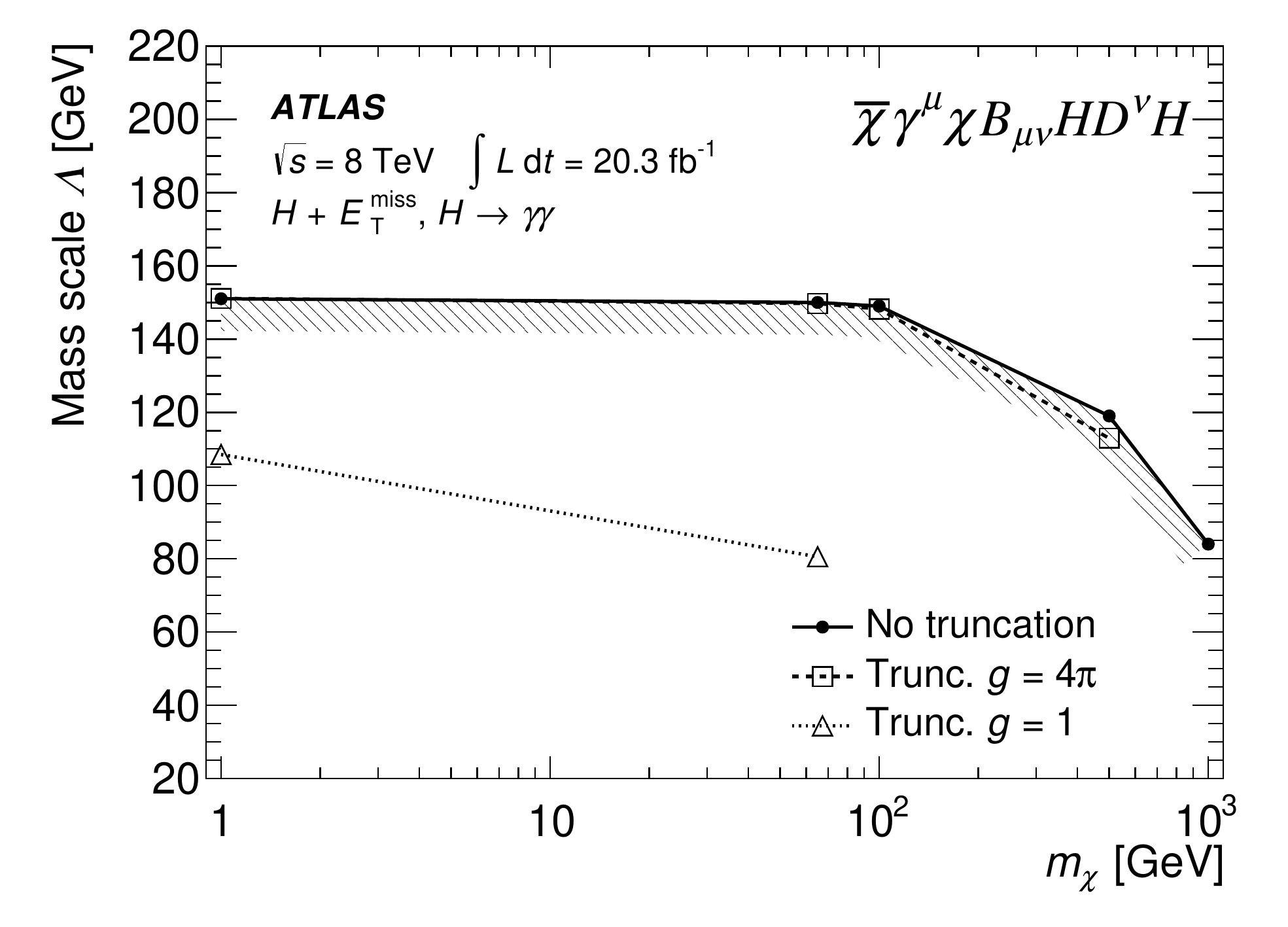}
        \includegraphics[width=0.495\textwidth]{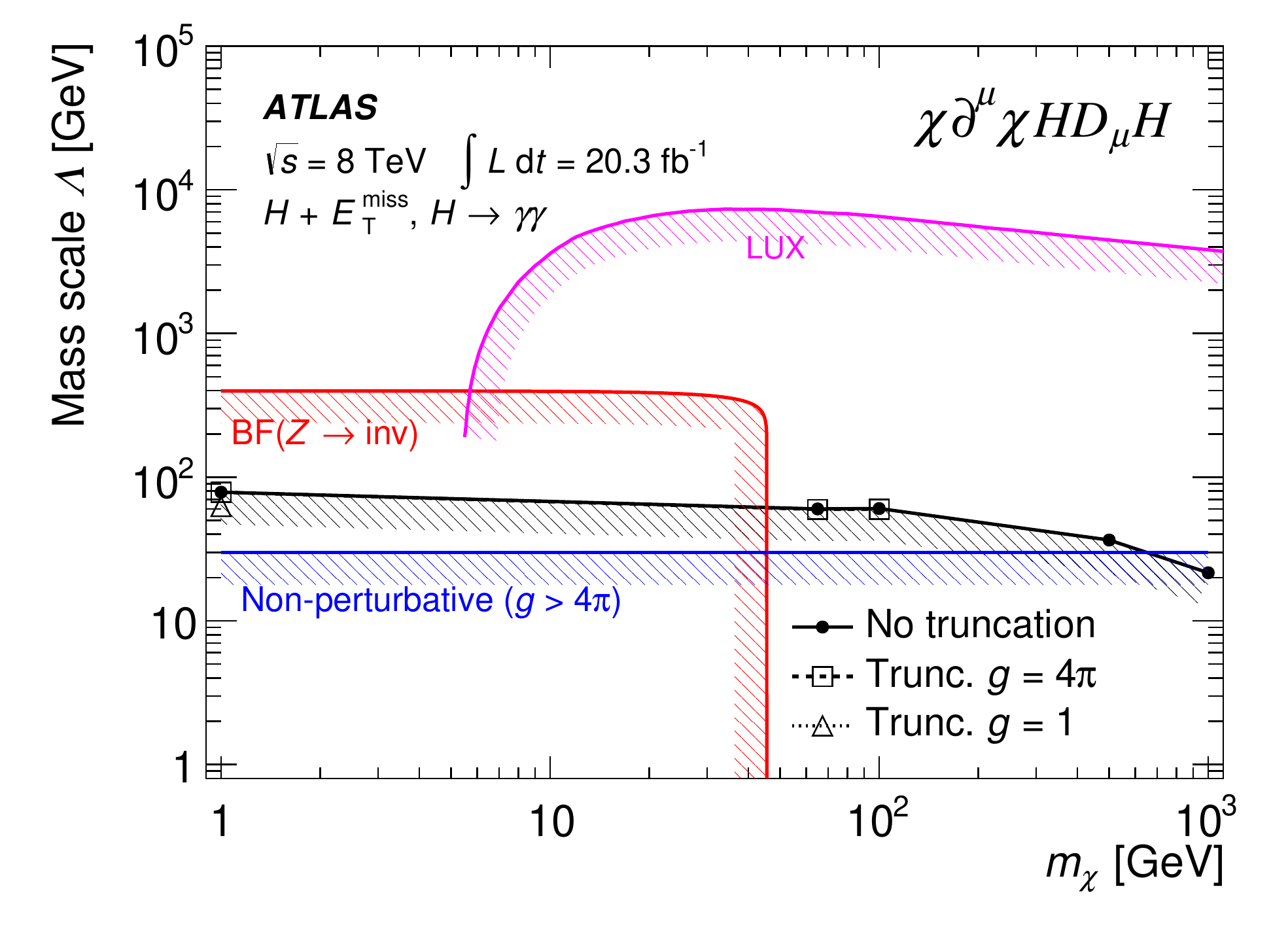}
\end{center}
\caption{Limits at 95\% CL on the mass scale $\Lambda$  as a function of the DM mass ($m_\chi$) for two of the four EFT models considered. Solid black lines are due to $H + \MET$ (this Letter); results where EFT truncation is applied are also shown, assuming coupling values $g=\sqrt{g_qg_\chi} = 1,4\pi$. The $g=4\pi$ case overlaps with the no-truncation result. The blue line indicates regions that fail the perturbativity requirement of $g<4\pi$, the red line indicates regions excluded by $Z$ boson limits~\cite{Beringer:1900zz} on the invisible branching fraction (BF), and the pink line indicates regions excluded by the LUX Collaboration~\cite{Akerib:2013tjd}.}
\label{fig:eft-limits}
\end{figure}

\begin{figure}
\begin{center}
        \includegraphics[width=0.495\textwidth]{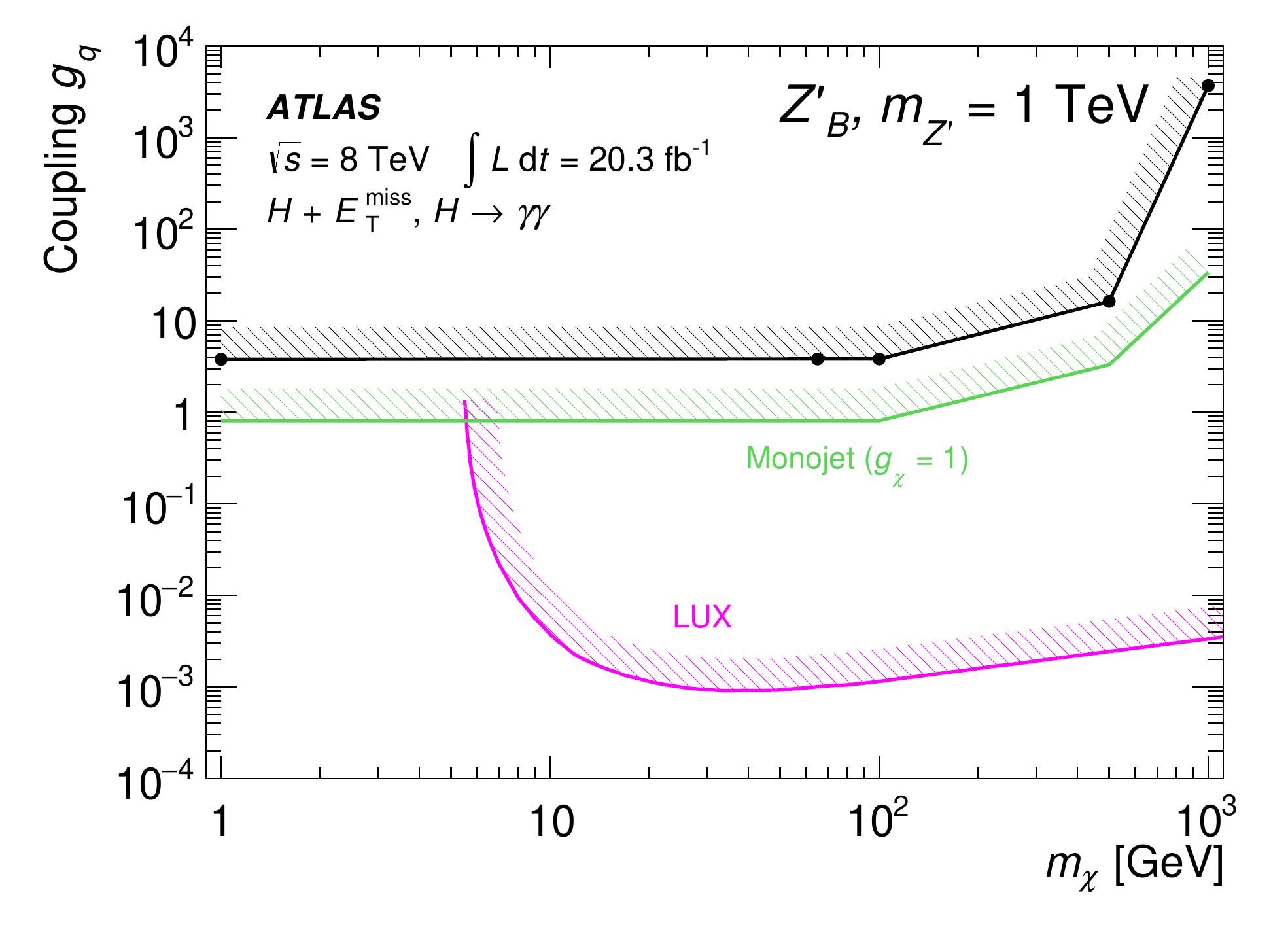}
        \includegraphics[width=0.495\textwidth]{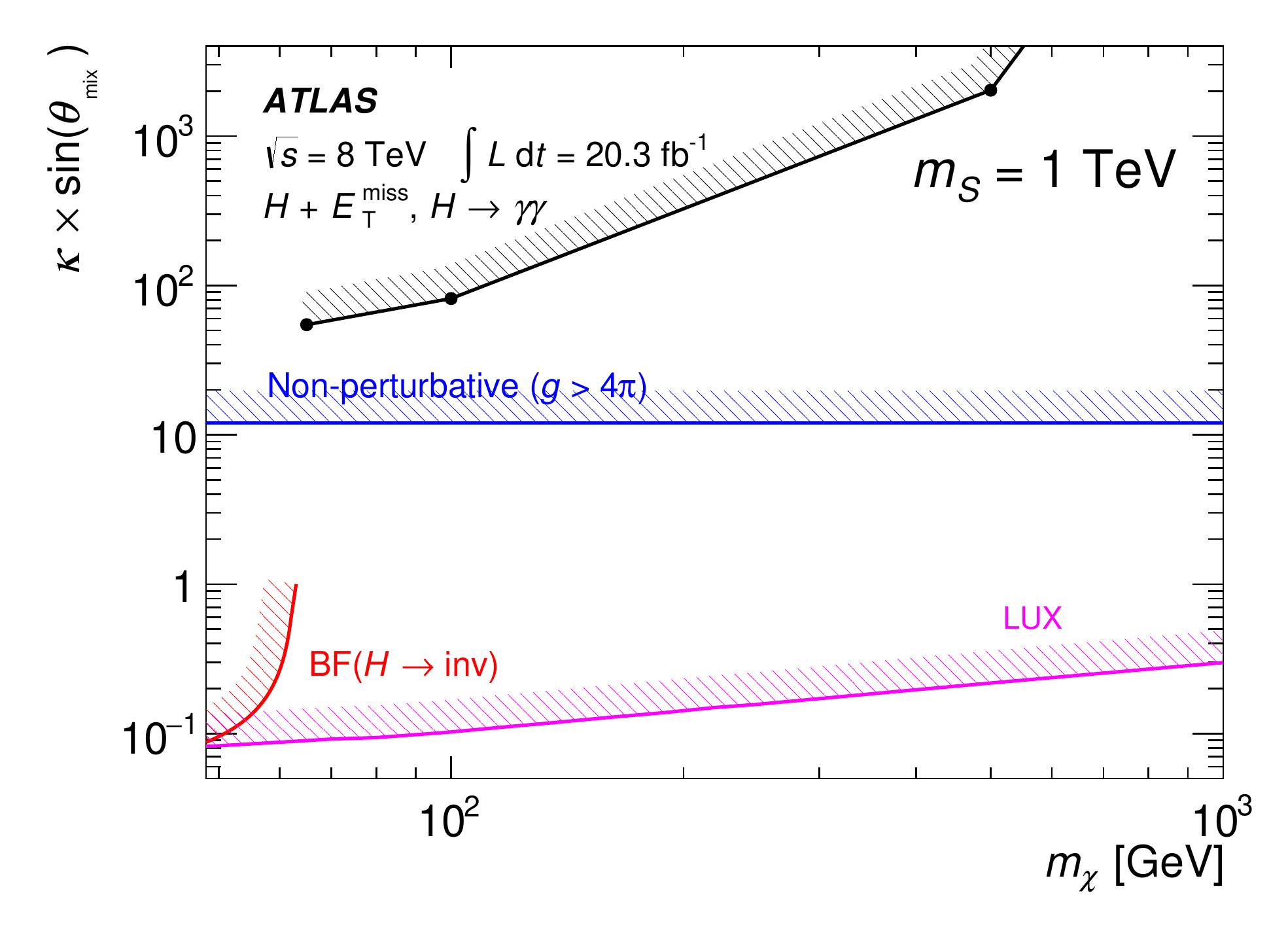}

\end{center}
\caption{ Limits on coupling parameters for simplified models with a heavy mediator with mass of 1 TeV. All constraint contours exclude larger couplings or mixing angles. Regions excluded due to perturbativity arguments are indicated; red, green and pink contours denote results from collider searches for invisible $H$ decays~\cite{Belanger:2013xza}, and monojet~\cite{Aad:2011xw} searches, and the LUX Collaboration~\cite{Akerib:2013tjd}, respectively.}
\label{fig:sm-limits}
\end{figure}

Limits on DM production are derived from the cross-section limits at a given DM mass $m_\chi$, and expressed as 95\% CL limits on the suppression scale $\Lambda$ or coupling parameter $\lambda$ for the effective field theory operators; see Fig.~\ref{fig:eft-limits} for limits for $\chi^\dagger \partial^\mu \chi  H^\dagger D_\mu H$ and $\bar \chi \gamma^\mu \chi  B_{\mu\nu} H^\dagger D^\nu H$ operators. For the lowest $m_\chi$ region not excluded by results from searches for invisible Higgs boson decays near $m_\chi=m_{H}/2$, values of $\Lambda$  up to 6, 60, and 150 GeV are excluded for the  $\bar{\chi}i\gamma_5\chi|H|^2$, $\chi^\dagger \partial^\mu \chi  H^\dagger D_\mu H$, and $\bar \chi \gamma^\mu \chi  B_{\mu\nu} H^\dagger D^\nu H$ operators, respectively; values of $\lambda$ above 25.6 are excluded for the $|\chi|^2|H|^2$ operator.  As discussed above, the effective field theory model becomes a poor approximation of an ultraviolet-complete model containing a heavy mediator $V$ when the momentum transferred in the interaction, $Q_{\textrm{tr}}$, is comparable to the mass of the intermediate state $m_V = \Lambda\sqrt{g_q g_\chi}$~\cite{Goodman:2010ku,Busoni:2013lha}, where $g_q$ and $g_\chi$ represent the coupling of $V$ to SM and DM particles, respectively. To give an indication of the impact of the unknown ultraviolet details of the theory, limits are computed in which only simulated events with $Q_{\textrm{tr}}=m_{\chi\chi} < m_V$ are retained; these limits are  shown for values of $\sqrt{g_qg_\chi}=1$ or $4\pi$ in Fig.~\ref{fig:eft-limits}. This procedure is referred to as truncation. In addition, limits are derived on coupling parameters for simplified models as shown in Fig.~\ref{fig:sm-limits}. For a vector-mediated model, limits are placed on the coupling $g_q$ of the mediator to quarks, assuming maximal coupling $g_\chi$ to dark matter. For the scalar-mediated model, limits are placed on the parameter $\kappa \times \sin(\theta_{\textrm{mix}})$, where $\sin(\theta_{\textrm{mix}})$ is the mixing angle between the scalar $S$ boson and the Higgs boson, and $\kappa$ is a scaling constant; however, current calculations~\cite{Carpenter:2013xra} of the $gg\rightarrow HS$ production mode may be overestimated due to approximations made in evaluating the top-quark loop.

In conclusion, a search for DM produced in association with a Higgs boson decaying to two photons has been conducted. Prior to these results, no bounds have been placed by collider experiments on the $H$+DM models discussed here.  In addition, upper limits are placed on the cross section of events with large missing transverse momentum and a Higgs boson.

We thank CERN for the very successful operation of the LHC, as well as the support staff from our institutions without whom ATLAS could not be operated efficiently.

We acknowledge the support of ANPCyT, Argentina; YerPhI, Armenia; ARC, Australia; BMWFW and FWF, Austria; ANAS, Azerbaijan; SSTC, Belarus; CNPq and FAPESP, Brazil; NSERC, NRC and CFI, Canada; CERN; CONICYT, Chile; CAS, MOST and NSFC, China; COLCIENCIAS, Colombia; MSMT CR, MPO CR and VSC CR, Czech Republic; DNRF, DNSRC and Lundbeck Foundation, Denmark; EPLANET, ERC and NSRF, European Union; IN2P3-CNRS, CEA-DSM/IRFU, France; GNSF, Georgia; BMBF, DFG, HGF, MPG and AvH Foundation, Germany; GSRT and NSRF, Greece; ISF, MINERVA, GIF, I-CORE and Benoziyo Center, Israel; INFN, Italy; MEXT and JSPS, Japan; CNRST, Morocco; FOM and NWO, Netherlands; BRF and RCN, Norway; MNiSW and NCN, Poland; GRICES and FCT, Portugal; MNE/IFA, Romania; MES of Russia and ROSATOM, Russian Federation; JINR; MSTD, Serbia; MSSR, Slovakia; ARRS and MIZŠ, Slovenia; DST/NRF, South Africa; MINECO, Spain; SRC and Wallenberg Foundation, Sweden; SER, SNSF and Cantons of Bern and Geneva, Switzerland; NSC, Taiwan; TAEK, Turkey; STFC, the Royal Society and Leverhulme Trust, United Kingdom; DOE and NSF, United States of America.
The crucial computing support from all WLCG partners is acknowledged gratefully, in particular from CERN and the ATLAS Tier-1 facilities at TRIUMF (Canada), NDGF (Denmark, Norway, Sweden), CC-IN2P3 (France), KIT/GridKA (Germany), INFN-CNAF (Italy), NL-T1 (Netherlands), PIC (Spain), ASGC (Taiwan), RAL (UK) and BNL (USA) and in the Tier-2 facilities worldwide.

\bibliographystyle{atlasBibStyleWoTitle}

\bibliography{monoh}

\providecommand{\href}[2]{#2}\begingroup\raggedright\begin{thebibliography}{10}

\bibitem{Bertone:2004pz}
G.~Bertone, D.~Hooper,  and J.~Silk,
  \href{http://dx.doi.org/10.1016/j.physrep.2004.08.031}{Phys. Rept. {\bfseries
  405} (2005) 279--390},
\href{http://arxiv.org/abs/hep-ph/0404175}{{\ttfamily arXiv:hep-ph/0404175
  [hep-ph]}}.

\bibitem{Jungman:1995df}
G.~Jungman, M.~Kamionkowski,  and K.~Griest,
  \href{http://dx.doi.org/10.1016/0370-1573(95)00058-5}{Phys. Rept. {\bfseries
  267} (1996) 195--373},
\href{http://arxiv.org/abs/hep-ph/9506380}{{\ttfamily arXiv:hep-ph/9506380
  [hep-ph]}}.

\bibitem{Scherrer:1985zt}
R.~J. Scherrer and M.~S. Turner,
\href{http://dx.doi.org/10.1103/PhysRevD.33.1585}{Phys. Rev. {\bfseries D33}
  (1986) 1585}.

\bibitem{coord}
ATLAS uses a right-handed coordinate system with its origin at the nominal
  interaction point (IP) in the center of the detector and the $z$-axis along
  the beam pipe. The $x$-axis points from the IP to the center of the LHC ring,
  and the $y$-axis points upward. Cylindrical coordinates ($r,\phi$) are used
  in the transverse plane, $\phi$ being the azimuthal angle around the beam
  pipe. The pseudorapidity is defined in terms of the polar $\theta$ angle as
  ${\eta = -\ln\left[\tan(\theta/2)\right]}$. The transverse energy is defined
  by $E_{\rm T} = E\sin{\theta}$.

\bibitem{Chatrchyan:2012me}
{CMS} Collaboration, \href{http://dx.doi.org/10.1007/JHEP09(2012)094}{JHEP
  {\bfseries 1209} (2012) 094},
\href{http://arxiv.org/abs/1206.5663}{{\ttfamily arXiv:1206.5663 [hep-ex]}}.

\bibitem{Aad:2011xw}
{ATLAS} Collaboration,
  \href{http://dx.doi.org/10.1016/j.physletb.2011.10.006}{Phys. Lett.
  {\bfseries B705} (2011) 294--312},
\href{http://arxiv.org/abs/1106.5327}{{\ttfamily arXiv:1106.5327 [hep-ex]}}.

\bibitem{Aad:2012fw}
{ATLAS} Collaboration,
  \href{http://dx.doi.org/10.1103/PhysRevLett.110.011802}{Phys. Rev. Lett.
  {\bfseries 110} (2013) 011802},
\href{http://arxiv.org/abs/1209.4625}{{\ttfamily arXiv:1209.4625 [hep-ex]}}.

\bibitem{Chatrchyan:2012tea}
{CMS} Collaboration,
  \href{http://dx.doi.org/10.1103/PhysRevLett.108.261803}{Phys. Rev. Lett.
  {\bfseries 108} (2012) 261803},
\href{http://arxiv.org/abs/1204.0821}{{\ttfamily arXiv:1204.0821 [hep-ex]}}.

\bibitem{Carpenter:2012rg}
L.~M. Carpenter, A.~Nelson, C.~Shimmin, T.~M.~P. Tait,  and D.~Whiteson,
  \href{http://dx.doi.org/10.1103/PhysRevD.87.074005}{Phys. Rev. {\bfseries
  D87} (2013) 074005},
\href{http://arxiv.org/abs/1212.3352}{{\ttfamily arXiv:1212.3352}}.

\bibitem{Aad:2014vka}
{ATLAS} Collaboration,
  \href{http://dx.doi.org/10.1103/PhysRevD.90.012004}{Phys. Rev. {\bfseries
  D90} (2014) 012004},
\href{http://arxiv.org/abs/1404.0051}{{\ttfamily arXiv:1404.0051 [hep-ex]}}.

\bibitem{Aad:2013oja}
{ATLAS} Collaboration,
  \href{http://dx.doi.org/10.1103/PhysRevLett.112.041802}{Phys. Rev. Lett.
  {\bfseries 112} (2014) 041802},
\href{http://arxiv.org/abs/1309.4017}{{\ttfamily arXiv:1309.4017 [hep-ex]}}.

\bibitem{Aad:2012tfa}
{ATLAS} Collaboration,
  \href{http://dx.doi.org/10.1016/j.physletb.2012.08.020}{Phys. Lett.
  {\bfseries B716} (2012) 1--29},
\href{http://arxiv.org/abs/1207.7214}{{\ttfamily arXiv:1207.7214 [hep-ex]}}.

\bibitem{Chatrchyan:2012ufa}
{CMS} Collaboration,
  \href{http://dx.doi.org/10.1016/j.physletb.2012.08.021}{Phys. Lett.
  {\bfseries B716} (2012) 30--61},
\href{http://arxiv.org/abs/1207.7235}{{\ttfamily arXiv:1207.7235 [hep-ex]}}.

\bibitem{Carpenter:2013xra}
L.~Carpenter {et~al.},
  \href{http://dx.doi.org/10.1103/PhysRevD.89.075017}{Phys. Rev. {\bfseries
  D89} (2014) 075017},
\href{http://arxiv.org/abs/1312.2592}{{\ttfamily arXiv:1312.2592 [hep-ph]}}.

\bibitem{Petrov:2013nia}
A.~A. Petrov and W.~Shepherd,
  \href{http://dx.doi.org/10.1016/j.physletb.2014.01.051}{Phys. Lett.
  {\bfseries B730} (2014) 178--183},
\href{http://arxiv.org/abs/1311.1511}{{\ttfamily arXiv:1311.1511 [hep-ph]}}.

\bibitem{Aad:2014iia}
{ATLAS} Collaboration,
  \href{http://dx.doi.org/10.1103/PhysRevLett.112.201802}{Phys. Rev. Lett.
  {\bfseries 112} (2014) 201802},
\href{http://arxiv.org/abs/1402.3244}{{\ttfamily arXiv:1402.3244 [hep-ex]}}.

\bibitem{Aad:2015uga}
{ATLAS} Collaboration, \href{http://arxiv.org/abs/1504.04324}{{\ttfamily
  arXiv:1504.04324 [hep-ex]}}.

\bibitem{Chatrchyan:2014tja}
{CMS} Collaboration,
  \href{http://dx.doi.org/10.1140/epjc/s10052-014-2980-6}{Eur.~Phys.~J.
  {\bfseries C74} (2014) 2980},
\href{http://arxiv.org/abs/1404.1344}{{\ttfamily arXiv:1404.1344 [hep-ex]}}.

\bibitem{ATLAS-CONF-2015-004}
{{ATLAS}} Collaboration, Report No. ATLAS-CONF-2015-004, (2015).
  \url{http://cds.cern.ch/record/2002121}.

\bibitem{Aad:2014lwa}
{ATLAS} Collaboration, \href{http://dx.doi.org/10.1007/JHEP09(2014)112}{JHEP
  {\bfseries 1409} (2014) 112},
\href{http://arxiv.org/abs/1407.4222}{{\ttfamily arXiv:1407.4222 [hep-ex]}}.

\bibitem{ATLAS}
{ATLAS} Collaboration,
  \href{http://dx.doi.org/doi:10.1088/1748-0221/3/08/S08003}{JINST {\bfseries
  3} (2008) S08003}.

\bibitem{Aad:2014nim}
{ATLAS} Collaboration,
  \href{http://dx.doi.org/10.1140/epjc/s10052-014-3071-4}{Eur. Phys. J.
  {\bfseries C74} (2014) 3071},
\href{http://arxiv.org/abs/1407.5063}{{\ttfamily arXiv:1407.5063 [hep-ex]}}.

\bibitem{Aad:2014eha}
{ATLAS} Collaboration,
  \href{http://dx.doi.org/10.1103/PhysRevD.90.112015}{Phys. Rev. {\bfseries
  D90} (2014) 112015},
\href{http://arxiv.org/abs/1408.7084}{{\ttfamily arXiv:1408.7084 [hep-ex]}}.

\bibitem{PhIsol}
{ATLAS} Collaboration,
  \href{http://dx.doi.org/10.1103/PhysRevD.83.052005}{Phys. Rev. D {\bfseries
  83} (2011) 052005},
\href{http://arxiv.org/abs/1012.4389}{{\ttfamily arXiv:1012.4389 [hep-ex]}}.

\bibitem{ATLAS-CONF-2013-082}
{{ATLAS}} Collaboration, Report No. ATLAS-CONF-2013-082, (2013).
  \url{http://cds.cern.ch/record/1570993}.

\bibitem{Aad:2012re}
{ATLAS} Collaboration, Eur. Phys. J. C {\bfseries 72} (2012) 1844,
  \href{http://arxiv.org/abs/1108.5602}{{\ttfamily arXiv:1108.5602 [hep-ex]}}.

\bibitem{Aad:2014bia}
{ATLAS} Collaboration,
  \href{http://dx.doi.org/10.1140/epjc/s10052-014-3190-y}{Eur. Phys. J.
  {\bfseries C75} (2015) 17},
\href{http://arxiv.org/abs/1406.0076}{{\ttfamily arXiv:1406.0076 [hep-ex]}}.

\bibitem{Alwall:2011uj}
J.~Alwall, M.~Herquet, F.~Maltoni, O.~Mattelaer,  and T.~Stelzer,
  \href{http://dx.doi.org/10.1007/JHEP06(2011)128}{JHEP {\bfseries 1106} (2011)
  128},
\href{http://arxiv.org/abs/1106.0522}{{\ttfamily arXiv:1106.0522 [hep-ph]}}.

\bibitem{Sjostrand:2007gs}
T.~Sj{\"o}strand, S.~Mrenna,  and P.~Z. Skands,
  \href{http://dx.doi.org/10.1016/j.cpc.2008.01.036}{Comput. Phys. Commun.
  {\bfseries 178} (2008) 852--867},
\href{http://arxiv.org/abs/0710.3820}{{\ttfamily arXiv:0710.3820 [hep-ph]}}.

\bibitem{ATL-PHYS-PUB-2011-008}
{{ATLAS}} Collaboration, Report No. ATL-PHYS-PUB-2011-008 (2011).
  \url{http://cds.cern.ch/record/1345343}.

\bibitem{Martin:2009iq}
A.~D. Martin, W.~J. Stirling, R.~S. Thorne,  and G.~Watt,
  \href{http://dx.doi.org/10.1140/epjc/s10052-009-1072-5}{Eur. Phys. J. C
  {\bfseries 63} (2009) 189--285},
\href{http://arxiv.org/abs/0901.0002}{{\ttfamily arXiv:0901.0002 [hep-ph]}}.

\bibitem{Pumplin:2002vw}
J.~Pumplin {et~al.}, J. High Energy Phys. {\bfseries 07} (2002) 012,
\href{http://arxiv.org/abs/hep-ph/0201195}{{\ttfamily arXiv:hep-ph/0201195
  [hep-ph]}}.

\bibitem{Han:1991ia}
T.~Han and S.~Willenbrock,
\href{http://dx.doi.org/10.1016/0370-2693(91)90572-8}{Phys. Lett. {\bfseries
  B273} (1991) 167--172}.

\bibitem{Brein:2003wg}
O.~Brein, A.~Djouadi,  and R.~Harlander,
  \href{http://dx.doi.org/10.1016/j.physletb.2003.10.112}{Phys. Lett.
  {\bfseries B579} (2004) 149--156},
\href{http://arxiv.org/abs/hep-ph/0307206}{{\ttfamily arXiv:hep-ph/0307206
  [hep-ph]}}.

\bibitem{Ciccolini:2003jy}
M.~L. Ciccolini, S.~Dittmaier,  and M.~Kr{\"a}mer,
  \href{http://dx.doi.org/10.1103/PhysRevD.68.073003}{Phys. Rev. {\bfseries
  D68} (2003) 073003},
\href{http://arxiv.org/abs/hep-ph/0306234}{{\ttfamily arXiv:hep-ph/0306234
  [hep-ph]}}.

\bibitem{:2010wqa}
{ATLAS} Collaboration,
  \href{http://dx.doi.org/10.1140/epjc/s10052-010-1429-9}{Eur. Phys. J. C
  {\bfseries 70} (2010) 823--874},
\href{http://arxiv.org/abs/1005.4568}{{\ttfamily arXiv:1005.4568
  [physics.ins-det]}}.

\bibitem{Agostinelli:2002hh}
{GEANT4} Collaboration, S.~Agostinelli {et~al.},
\href{http://dx.doi.org/10.1016/S0168-9002(03)01368-8}{Nucl. Instrum. Methods
  Phys. Res., Sect. A {\bfseries 506} (2003) 250--303}.

\bibitem{Oreglia:1980cs}
M.~Oreglia,
Ph.D. Thesis (Stanford Univ.) (1980).

\bibitem{Aad:2013ucp}
{ATLAS} Collaboration,
  \href{http://dx.doi.org/10.1140/epjc/s10052-013-2518-3}{Eur. Phys. J.
  {\bfseries C73} (2013) 2518},
\href{http://arxiv.org/abs/1302.4393}{{\ttfamily arXiv:1302.4393 [hep-ex]}}.

\bibitem{Lai:2010vv}
H.-L. Lai {et~al.}, \href{http://dx.doi.org/10.1103/PhysRevD.82.074024}{Phys.
  Rev. D {\bfseries 82} (2010) 074024},
\href{http://arxiv.org/abs/1007.2241}{{\ttfamily arXiv:1007.2241 [hep-ph]}}.

\bibitem{Botje01}
M.~Botje {et~al.}, \href{http://arxiv.org/abs/1101.0538}{{\ttfamily
  arXiv:1101.0538 [hep-ph]}}.

\bibitem{Ball:2011mu}
{{NNPDF}} Collaboration, R.~D. Ball {et~al.},
  \href{http://dx.doi.org/10.1016/j.nuclphysb.2011.03.021}{Nucl. Phys.
  {\bfseries B849} (2011) 296--363},
\href{http://arxiv.org/abs/1101.1300}{{\ttfamily arXiv:1101.1300 [hep-ph]}}.

\bibitem{Djouadi:1997yw}
A.~Djouadi, J.~Kalinowski,  and M.~Spira,
  \href{http://dx.doi.org/10.1016/S0010-4655(97)00123-9}{Comput. Phys. Commun.
  {\bfseries 108} (1998) 56--74},
\href{http://arxiv.org/abs/hep-ph/9704448}{{\ttfamily arXiv:hep-ph/9704448
  [hep-ph]}}.

\bibitem{Actis:2008ts}
S.~Actis, G.~Passarino, C.~Sturm,  and S.~Uccirati,
  \href{http://dx.doi.org/10.1016/j.nuclphysb.2008.11.024}{Nucl. Phys.
  {\bfseries B811} (2009) 182--273},
\href{http://arxiv.org/abs/0809.3667}{{\ttfamily arXiv:0809.3667 [hep-ph]}}.

\bibitem{Dittmaier:2011ti}
{LHC Higgs Cross Section Working Group, S. Dittmaier, C. Mariotti, G.
  Passarino, and R. Tanaka (Eds.)},
\href{http://arxiv.org/abs/1101.0593}{{\ttfamily arXiv:1101.0593 [hep-ph]}}.

\bibitem{Dittmaier:2012vm}
{LHC Higgs Cross Section Working Group, S. Dittmaier, C. Mariotti, G.
  Passarino, and R. Tanaka (Eds.)},
\href{http://arxiv.org/abs/1201.3084}{{\ttfamily arXiv:1201.3084 [hep-ph]}}.

\bibitem{Cowan:2010js}
G.~Cowan, K.~Cranmer, E.~Gross,  and O.~Vitells,
  \href{http://dx.doi.org/10.1140/epjc/s10052-011-1554-0}{Eur. Phys. J.
  {\bfseries C71} (2011) 1554},
\href{http://arxiv.org/abs/1007.1727}{{\ttfamily arXiv:1007.1727
  [physics.data-an]}}.

\bibitem{Read:2002hq}
A.~L. Read,
\href{http://dx.doi.org/10.1088/0954-3899/28/10/313}{J. Phys. {\bfseries G28}
  (2002) 2693--2704}.

\bibitem{Junk:1999kv}
T.~Junk, \href{http://dx.doi.org/10.1016/S0168-9002(99)00498-2}{Nucl. Instrum.
  Meth. {\bfseries A434} (1999) 435--443},
\href{http://arxiv.org/abs/hep-ex/9902006}{{\ttfamily arXiv:hep-ex/9902006
  [hep-ex]}}.

\bibitem{Cranmer:2013hia}
K.~Cranmer, S.~Kreiss, D.~Lopez-Val,  and T.~Plehn,
  \href{http://dx.doi.org/10.1103/PhysRevD.91.054032}{Phys.Rev. {\bfseries D91}
  (2015) 054032},
\href{http://arxiv.org/abs/1401.0080}{{\ttfamily arXiv:1401.0080 [hep-ph]}}.

\bibitem{Beringer:1900zz}
{Particle Data Group, J. Beringer et al.},
\href{http://dx.doi.org/10.1103/PhysRevD.86.010001}{Phys. Rev. {\bfseries D86}
  (2012) 010001}.

\bibitem{Akerib:2013tjd}
{LUX} Collaboration, D.~S. Akerib {et~al.},
  \href{http://dx.doi.org/10.1103/PhysRevLett.112.091303}{Phys. Rev. Lett.
  {\bfseries 112} (2014) 091303},
\href{http://arxiv.org/abs/1310.8214}{{\ttfamily arXiv:1310.8214
  [astro-ph.CO]}}.

\bibitem{Belanger:2013xza}
G.~B{\'e}langer, B.~Dumont, U.~Ellwanger, J.~F. Gunion,  and S.~Kraml,
  \href{http://dx.doi.org/10.1103/PhysRevD.88.075008}{Phys. Rev. {\bfseries
  D88} (2013) 075008},
\href{http://arxiv.org/abs/1306.2941}{{\ttfamily arXiv:1306.2941 [hep-ph]}}.

\bibitem{Goodman:2010ku}
J.~Goodman {et~al.}, \href{http://dx.doi.org/10.1103/PhysRevD.82.116010}{Phys.
  Rev. {\bfseries D82} (2010) 116010},
\href{http://arxiv.org/abs/1008.1783}{{\ttfamily arXiv:1008.1783 [hep-ph]}}.

\bibitem{Busoni:2013lha}
G.~Busoni, A.~De~Simone, E.~Morgante,  and A.~Riotto,
  \href{http://dx.doi.org/10.1016/j.physletb.2013.11.069}{Phys. Lett. B
  {\bfseries 728} (2014) 412--421},
\href{http://arxiv.org/abs/1307.2253}{{\ttfamily arXiv:1307.2253 [hep-ph]}}.

\end{thebibliography}\endgroup

\clearpage 
\begin{flushleft}
{\Large The ATLAS Collaboration}

\bigskip

G.~Aad$^{\rm 85}$,
B.~Abbott$^{\rm 113}$,
J.~Abdallah$^{\rm 151}$,
O.~Abdinov$^{\rm 11}$,
R.~Aben$^{\rm 107}$,
M.~Abolins$^{\rm 90}$,
O.S.~AbouZeid$^{\rm 158}$,
H.~Abramowicz$^{\rm 153}$,
H.~Abreu$^{\rm 152}$,
R.~Abreu$^{\rm 30}$
% ...
\bigskip
\\
% ...
$^{11}$ Institute of Physics, Azerbaijan Academy of Sciences, Baku, Azerbaijan\\
% ...
$^{30}$ CERN, Geneva, Switzerland\\
% ...
$^{85}$ CPPM, Aix-Marseille Universit{\'e} and CNRS/IN2P3, Marseille, France\\
% ...
$^{90}$ Department of Physics and Astronomy, Michigan State University, East Lansing MI, United States of America\\
% ...
$^{107}$ Nikhef National Institute for Subatomic Physics and University of Amsterdam, Amsterdam, Netherlands\\
% ...
$^{113}$ Homer L. Dodge Department of Physics and Astronomy, University of Oklahoma, Norman OK, United States of America\\
% ...
$^{151}$ Institute of Physics, Academia Sinica, Taipei, Taiwan\\
$^{152}$ Department of Physics, Technion: Israel Institute of Technology, Haifa, Israel\\
$^{153}$ Raymond and Beverly Sackler School of Physics and Astronomy, Tel Aviv University, Tel Aviv, Israel\\
% ...
$^{158}$ Department of Physics, University of Toronto, Toronto ON, Canada\\
% ...
\end{flushleft}

\end{document}